\documentclass[conference]{IEEEtran}
\IEEEoverridecommandlockouts
\usepackage{cite}
\usepackage{amsmath,amssymb,amsfonts}
\usepackage{algorithmic}
\usepackage{graphicx}
\usepackage{textcomp}
\usepackage{xcolor}

\usepackage{siunitx}
\usepackage{lineno}

 \usepackage{float} 
	\usepackage{graphicx,rotating,booktabs}
    \usepackage{graphics}
	\usepackage{amsmath}
	\usepackage{pdfpages}
	\usepackage[resetlabels]{multibib}
    \usepackage{mathtools}
    \usepackage{setspace}
    \usepackage{rotating}
\newcites{pp}{References}
\usepackage{lineno,hyperref}
\usepackage[verbose]{placeins}
\usepackage{blindtext}
\usepackage{amsmath}
\usepackage{tikz}

\def\BibTeX{{\rm B\kern-.05em{\sc i\kern-.025em b}\kern-.08em
    T\kern-.1667em\lower.7ex\hbox{E}\kern-.125emX}}
\begin{document}

\title{A Comprehensive Analysis of Correlated Source Compression Using Edge Computing in Distributed Systems}
\author{\IEEEauthorblockN{Benjamin Rosen}
\IEEEauthorblockA{\textit{School of Electrical and Information Engineering} \\
\textit{University of the Witwatersrand}\\
Johannesburg, South Africa \\
benjamin.rosen@students.wits.ac.za}
\and
\IEEEauthorblockN{Shane House}
\IEEEauthorblockA{\textit{School of Electrical and Information Engineering} \\
\textit{University of the Witwatersrand}\\
Johannesburg, South Africa \\
shane.house@students.wits.ac.za}
\and
\IEEEauthorblockN{Shamin Achari}
\IEEEauthorblockA{\textit{School of Electrical and Information Engineering} \\
\textit{University of the Witwatersrand}\\
Johannesburg, South Africa \\
Shamin.Achari@wits.ac.za}
\and
\IEEEauthorblockN{Ling Cheng}
\IEEEauthorblockA{\textit{School of Electrical and Information Engineering} \\
\textit{University of the Witwatersrand}\\
Johannesburg, South Africa \\
Ling.Cheng@wits.ac.za}
}

\maketitle

\begin{abstract}
This paper examines the theory pertaining to lossless compression of correlated sources located at the edge of a network. Importantly, communication between nodes is prohibited. In particular, a method that combines correlated source coding and matrix partitioning is explained. This technique is then made more flexible, by restricting the method to operate on two distinct groups of nodes. As a result, this new method allows for more freedom in compression performance, with consequent trade-off in node integrity validation. Specifically, it provides 2-3 times the compression savings when using a Hamming(7,4) with 4 nodes. It also decreases the complexity with regard to managing the nodes as they join/leave the network, while retaining the range within which the information can be losslessly decoded.

\end{abstract}


\section{Introduction}
\IEEEPARstart{E}{dge} computing is a form of a distributed system where the processing is generally performed by the sensing nodes at the ends of a network. This tends to reduce the processing load on the central server as well as reduce the amount of information that is transmitted back and forth between nodes and servers.
In the current age of IoT (Internet of Things) where everyday items such as scales, lights, televisions, toothbrushes and even kettles all connect to the internet, it has become ever more important for good coding and compression techniques to ensure effective usage of the limited bandwidth.

In their pioneering work on correlated source coding, Slepian and
Wolf \cite{1055037} first showed how distributed sources can compress
data and decode at a central point in a lossless manner. They present
various scenarios and configurations that demonstrate this theory,
the most relevant of which is source coding without communication
between sources. Additionally, they graph the maximum rate limits for
this scheme. Their work was extended numerous times, most notably
by Pradhan \emph{et al.} \cite{1281474,755665,838176},
who propose methods to achieve any arbitrary point in the Slepian-Wolf
bound. Xiong \emph{et al. }\cite{1328091} describes
how this method could be applied to sensor networks (a focus of this
paper) and use list decoding to implement this. Stankovi\'{c} \emph{et
al.} \cite{1281475} demonstrate a more realisable
form of the encoding/decoding scheme, which is used in this paper
to create the Flexible Grouping method. S. Choi \cite{4815077} uses
graph techniques to compress data in a distributed system. However,
the method is lossy and requires the use of MAC protocols, which fall
outside the scope of this work. A. D. Liveris \emph{et al.} \cite{1033198} present an interesting technique to compress images,
but use turbo-codes to do so, thus falling outside the aim of this
work to use matrix partitioning with linear block codes.

This paper develops and analyses such compression schemes which make use of correlated source coding. These schemes will be used in the context of a distributed system and in particular, they will utilise the power of edge computing by shifting the complexity and processing to the sensor nodes at the edge of the network. Specifically, this paper improves the scheme outlined by Stankovi\'{c} \emph{et
al.} \cite{1281475} in terms of compression efficiency. By creating a more space-saving compression scheme, the overall throughput of the system can be increased, leading to more information at the central processing node. This in turn will result in an Information Fusion system that is more intelligent, provides a greater insight into the measured event and is therefore able to respond faster to event phenomena.

Section \ref{sec:background} of this article outlines the background and theories in this field. Section \ref{sec:schemes} provides an overview of the system model and  describes the two different compression algorithms. Section \ref{sec:metrics} discusses the various metrics used to analyse the performance of the schemes. The results and analysis are discussed in Section \ref{sec:performance} and finally, the paper is concluded in Section \ref{sec:conclusion}.

\section{Background\label{sec:background}}
Consider two encoders $x$ and $y$, with corresponding correlated random variables $X$ and $Y$. If the encoders are allowed to communicate amongst themselves as well as a common decoder, the rate region in which they can operate without loss of information is bounded by $R_{x}+R_{y}\geq H(X,Y)$. Slepian and Wolf \cite{1055037} showed, somewhat surprisingly, that this rate region is still achievable even if the encoders are forbidden from sharing information with one another, provided that it is further bounded by Equations \eqref{eq:Rx} and \eqref{eq:Ry}:
\begin{equation}
R_{x}\geq H(X|Y)\label{eq:Rx}
\end{equation}
\begin{equation}
R_{y}\geq H(Y|X)\label{eq:Ry}
\end{equation}
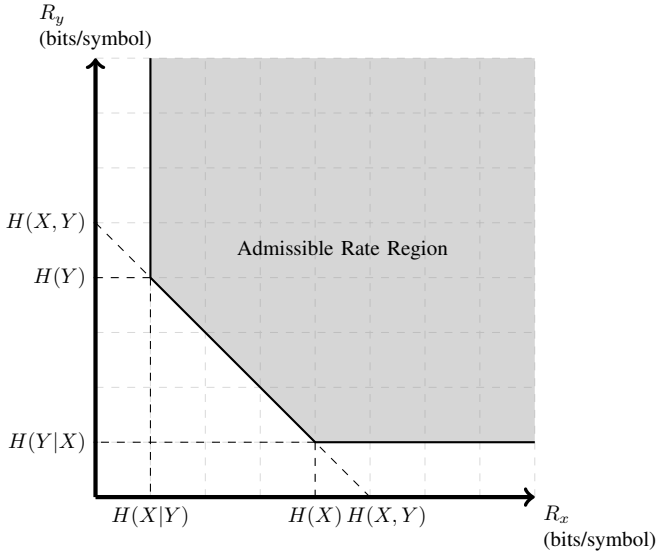
\begin{figure}
\begin{tikzpicture} [scale=0.73, every node/.style={scale=0.8}]
\draw[help lines, color=gray!30, dashed] (0,0) grid (8,8);
\draw[->,ultra thick] (0,0)--(8,0) node[below right, align=left]{$R_x$\\(bits/symbol)}; 
\draw[->,ultra thick] (0,0)--(0,8) node[above, align=left]{$R_y$\\(bits/symbol)};

\fill[white!60!black,fill opacity=0.4] (1,8)--(1,4)--(4,1)--(8,1)--(8,8)--cycle;

\draw[dashed] (0,5)--(5,0);
\draw[dashed] (4,1)--(0,1);
\draw[dashed] (1,4)--(1,0);
\draw[dashed] (4,1)--(4,0);
\draw[dashed] (1,4)--(0,4);

\draw[thick] (1,8)--(1,4)--(4,1)--(8,1);
\draw (1 cm,1pt) -- (1 cm,-1pt) node[anchor=north] {$H(X|Y)$};
\draw (4 cm,1pt) -- (4 cm,-1pt) node[anchor=north] {$H(X)$};
\draw (5.3 cm,1pt) -- (5.3 cm,-1pt) node[anchor=north] {$H(X,Y)$};

\draw (1pt,1 cm) -- (-1pt,1 cm) node[anchor=east] {$H(Y|X)$};
\draw (1pt,4 cm) -- (-1pt,4 cm) node[anchor=east] {$H(Y)$};
\draw (1pt,5 cm) -- (-1pt,5 cm) node[anchor=east] {$H(X,Y)$};
\node at (4.5,4.5) {Admissible Rate Region};

\end{tikzpicture}
\caption{Slepian-Wolf admissible rate region for two correlated sources without
information sharing.\label{fig:Slepian-Wolf-admissible-rate-no-info-sharing}}
\end{figure}
Figure \ref{fig:Slepian-Wolf-admissible-rate-no-info-sharing} shows this region, where the rates of $x$ and $y$ are given in bits per symbol. As a result, a lower rate means improved performance.

When looking at the edge of the bound, two regions become apparent. One is the corner point region (shown by the points $(H(X|Y),H(Y))$ and  $(H(X),H(Y|X))$ in Figure \ref{fig:Slepian-Wolf-admissible-rate-no-info-sharing}), while the other is the line $R_x+R_y= H(X,Y)$. The former is known as the asymmetric region, where one node sends full information and the other sends information that is fully compressed, while the latter is called the symmetric region, or where each node's information is partially compressed. The symmetric region is more desirable, as it spreads the load of the network amongst the nodes. Pradhan and Ramchandran \cite{838176} proposed that error correcting codes (ECC) could be used to achieve the Slepian-Wolf region. The basic idea is to model the correlation between nodes as a "virtual" channel. Thus, an ECC that is powerful enough to correct "errors" (i.e. the part of the information that is not correlated) resulting from this channel is a good source code for the Slepian-Wolf case. To achieve an arbitrary point on the admissible rate region, Matrix Partitioning (MP) is performed on the parity check matrix of the ECC, such that the rows are distributed amongst the nodes and they will send partial syndromes to the joint decoder. The decoder is then able to find the codewords that correspond to these syndromes and recover the nodes' information. Their idea was formalised by Stankovi\'{c} \emph{et
al.} \cite{1281475}, whose exact method is elaborated upon in Section \ref{subsec:DG}.

\section{Compression Model}\label{sec:schemes}

\begin{figure}[H]
\begin{centering}
\includegraphics[width=1\columnwidth]{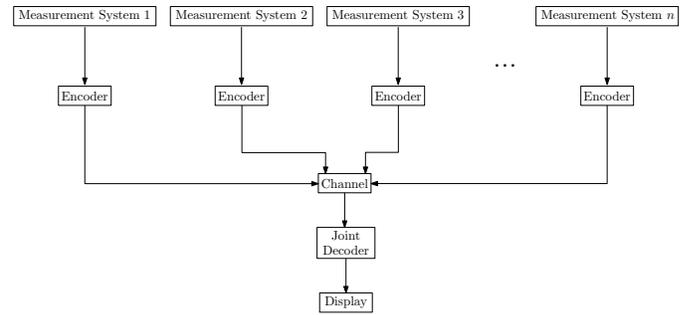}
\par\end{centering}
\caption{Overview of the system model with edge sensors and joint decoder server.\label{fig:overview}}
\end{figure}

The edge computing model makes use of multiple nodes and sensors which are connected at the end of the network. These edge nodes essentially measure, collect and process data before individually transmitting this data to the central server. The unique data from the respective nodes are then combined at the server which acts as the decoder by performing the decompression algorithm on the received data. The server also has the task of displaying the data to the end-user. Figure \ref{fig:overview} illustrates the overview of the system model described.

This paper focuses on two compression algorithms. These are the  Disjoint Grouping (DG) and Flexible Grouping (FG) methods of compression. Both these algorithms make use of MP and ECC. 

\subsection{Disjoint Grouping (DG)}\label{subsec:DG}
\begin{figure}
\begin{centering}
\includegraphics[width=1\columnwidth]{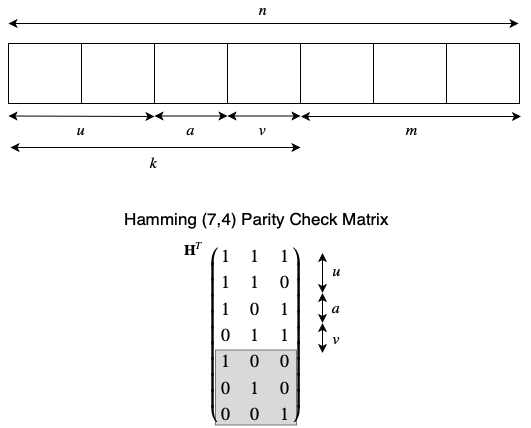}
\par\end{centering}
\caption{An example of the message composition when using the Hamming(7,4) code.\label{fig:MP-Comp}}

\end{figure}

The DG algorithm is first described in \cite{1281475} and will be used as a benchmark for the FG algorithm. Each node is assigned a set of sequential, non-overlapping rows, \textit{a}, from the \textbf{ $\mathbf{H}^{\mathit{T}}$} matrix. This is essentially the partitioning of the matrix, as \textit{a} determines not only which rows of the \textbf{ $\mathbf{H}^{\mathit{T}}$} matrix, but also the number of rows that each node is responsible for. Extra information, in the form of number of rows before \textit{a}, defined by \textit{u}, and number of rows after \textit{a}, defined by \textit{v}, are also required.  For each node, $u+a+v = k$.  Figure \ref{fig:MP-Comp} shows one possible combination of the above variables for a Hamming(7,4) code, with $u=2$, $a=1$ and $v=1$. Of importance is the fact that \textit{n}, \textit{k} and \textit{$m=n-k$} are all fixed by the code employed in the scheme, meaning that a suitable \textit{n} should be determined and cannot be arbitrarily assigned. Furthermore, the choices of \textit{u}, \textit{a} and \textit{v} are constrained to use only the systematic part of \textbf{$\mathbf{H}^{\mathit{T}}$} (i.e. the parts in Figure \ref{fig:MP-Comp} not highlighted in grey). 

To compress the message, each encoder takes the selection of bits from
its respective message represented by $a_i$, and multiplies this by the
corresponding rows in \textbf{$\mathbf{H}^{\mathit{T}}$}. The result will be of length \textit{m}, and is added to the final \textit{m} bits of the message, defined as $\overline{m}$. Each node will transmit the message $t$ which is original message where the last m bits are replaced by $\overline{m}$ and omitting the bits represented by $a$.

In order for this method to work, all rows in
the parity check matrix must be divided amongst all the nodes, with no overlapping rows. Thus, every
successive node receives the next portion of the matrix. For example, the first node could have a \textit{u }of zero and an \textit{a }of
two. The next node's \textit{u }must be equal to three, so as not
to overlap with the previous node's rows. The rest of the nodes would be set up in this fashion.

At the decoder, the messages from all nodes
is first collected. They must then be padded with zeros which are inserted
in the position/s occupied by the bits in \textit{a}. The next step
is to perform the grouped decoding method \cite{1281475}. This can
be summarised as follows:
\begin{itemize}
\item XOR two nodes' values together (producing \textit{c}).
\item Correct this result, using the selected ECC decoding method
(resulting in \textit{C}).
\item Replace the zero-padded values of each node with its corresponding values in $C$.
\item For each of the two nodes, multiply the bits in the $a_i$ position
and add the result to the original node's $\overline{m}$ value.
\end{itemize}
The algorithm returns the tentative decoded value for both nodes.
This method must then be repeated with every unique combination of
nodes (order is not important) in a pairwise manner. Thereafter, a vote decode ensues,
in which the most common decoded value for each node is selected as
the correct version.

\subsection{Flexible Grouping (FG)}

\begin{figure}[H]
\begin{centering}
\includegraphics[width=1\columnwidth]{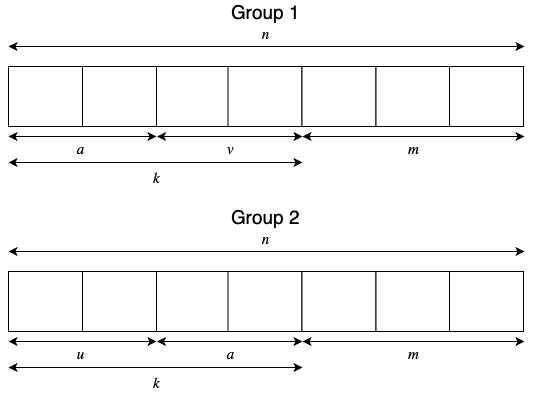}
\par\end{centering}
\caption{Example of the message construction for the Flexible Grouping method.\label{fig:FG-msg}}
\end{figure}

A unique scheme can be obtained from the DG method by constraining
the system to use only two groups and allowing multiple nodes to share
the same partition of the \textbf{$\mathbf{H}^{\mathit{T}}$} matrix.
Thus, all parameters are the same as the above method, with the additions
being the number of nodes in each group, $N_{g_{1}}$ and $N_{g_{2}}$, and how the rows are split
between the groups,  $r_{g_{1}}$ and $r_{g_{2}}$. This greatly simplifies the message construction,
which takes on two forms, as demonstrated in Figure \ref{fig:FG-msg}, which shows an equal row partition.
The decompression process uses the same padding and grouped decoding
method as in Section \ref{subsec:DG}, except that each node needs to be
decoded only once, without the vote decode. Another qualification is that the decoding must use nodes from different groups but does not have to include all possible combinations as required by the DG method.

\section{Analysis Metrics \label{sec:metrics}}
As with any system, a set of metrics needs to be established to quantify the performance of the various algorithms and parameters. While Bit Error Rate (BER) is generally used to characterise performance, here unique metrics are employed to rate the system. This research makes use of Compression Space Saving ($CSS$), lossless decoding range, information hiding/ data validation, decoding complexity, network management complexity and lastly node fault tolerance to quantify the effectiveness of the compression schemes. These items are discussed in detail below.

\subsection{Compression Space Saving (CSS)}
Compression Space Saving ($CSS$) describes the percentage of
the message (and therefore bandwidth) that is saved when performing
a particular compression. The basic governing equation of  is shown in Equation \eqref{eq:baseCSS}.

\begin{equation}
\label{eq:baseCSS}
    CSS = 1 - \frac{\text{compressed size}}{\text{uncompressed size}}
\end{equation}

\subsection{Lossless Decoding Range}\label{sec:LDR}
\begin{figure}
\begin{centering}
\includegraphics[width=1\columnwidth]{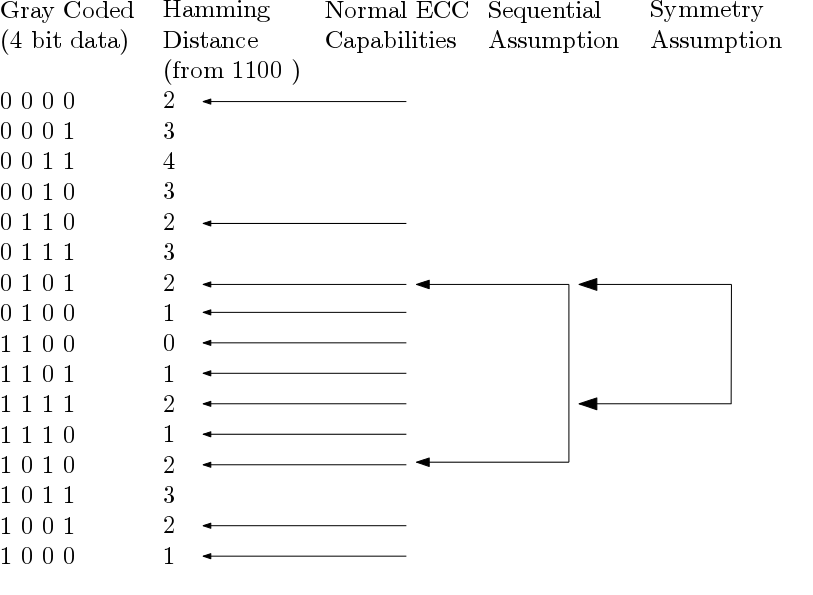}
\par\end{centering}
\caption{An example of the message composition when using the Hamming(7,4) code.\label{fig:ranges}}
\end{figure}

In general, error-correcting capabilities rely on the Hamming distance to determine the effectiveness of a coding scheme. This Hamming distance offers a good theoretical limit to the number of errors that may be corrected but it does not always reflect the true capabilities in a practical system. The Hamming distance provides a maximum of $2t+1$ errors that can theoretically be corrected. Since the proposed system is based on edge computing, various assumptions can be made to ensure a more practical measure is determined. The first assumption is that the data is sequential and follows a natural counting order which is inherent in real-world applications. This limits the number of errors that can be corrected as the corresponding erroneous codewords would need to be adjacent to the actual codeword. As such the next logical step is to use Gray coded data to ensure adjacent codewords only differ by at most one bit. The second assumption is that all sensor nodes are ideally measuring correlated conditions, and thus correlated quantities, so as one sensor's reading increases, all other sensors will also detect increased readings. These values may not be similar in quantity but are all proportional in relation to one another. The last assumption for the system is that the error range is symmetric around a given codeword. This assumption allows the system to have a working error correction range which is independent of the values measured by the nodes. All these assumptions effectively reduce the number of errors the system can correct to a smaller range which is called the lossless decoding range. This describes a guaranteed number of errors the system can correct. Figure \ref{fig:ranges} better illustrates the lossless decoding range for a $t=2$ error correction code and 4 bit Gray coded data. Figure \ref{fig:ranges} also shows the effects of each assumption mentioned above in reducing the range of errors corrected.

\subsection{Security and Node Integrity}
Security is provided in two layers in the proposed system. The first is based on the fact that data is compressed and thus information is left out of the transmitted data. Secondly, security is added by the fact that multiple nodes' data are required to get the full information as each node only transmits partial information.
Since the network is a distributed system, it can inherently make use of blockchain properties, where all the nodes perform calculations on a single transaction to prove integrity. The proposed system, however, uses correlation to couple the nodes and thus decoding takes place by making use of a majority vote where each node decodes a value based on every other node in the system. Since the final value used is based on a majority vote, at least half of the nodes' data will need to be altered in order to cause a change in the system. 

\subsection{Network Management Complexity}
This is the workload on the server when a new node is added to the system or an existing node is removed. This workload includes all the messages between the nodes and server to reorganise and restructure the system. Robustness is also taken into account in this metric as the network management complexity is additionally determined by how many nodes need to be in proper working condition for the system to still be fully functional. 

\section{Performance and Results}\label{sec:performance}

\subsection{Compression Space Saving}\label{subsec:CSS}
For the DG method with $N$ nodes, the $CSS$ is given by Equation \eqref{eqn:cssdg}, where $k$ and $n$ are defined by the ECC used and $k/n$ is the code rate.

\begin{equation}
CSS=\frac{k}{nN} , N<k
\label{eqn:cssdg}
\end{equation}
This implies that the compression worsens as more nodes are added to the scheme, and this will always be less than or equal the code rate.

For the FG method, the $CSS$ is defined by Equation \eqref{eqn:cssfg}.
\begin{equation}
CSS=\frac{N_{g_{1}}r_{g_{1}}+N_{g_{2}}r_{g_{2}}}{Nn}=\frac{r_{g_{1}}(N_{g_{1}}-N_{g_{2}})+N_{g_{2}}k}{Nn}\label{eqn:cssfg}
\end{equation}

Where: \medskip

\begin{tabular}{lll}
$N_{g_{x}}$  & = number of nodes in Group $x$ & \tabularnewline
$r_{g_{x}}$  & = number of rows in Group $x$ & \tabularnewline
$N_{g_{1}}+N_{g_{2}}$ & = $N$ & \tabularnewline
$r_{g_{1}}+r_{g_{2}}$ & = $k$ & \tabularnewline\\
\end{tabular}
\\
Depending on how the scheme is set up, different $CSS$ values will
be obtained. First, consider the case where the nodes and rows are
evenly split. Thus, $N_{g_{1}}=N_{g_{2}}$ and $r_{g_{1}}=r_{g_{2}}$. Equation \eqref{eqn:cssfg}
then reduces to:

\begin{equation}
CSS=\frac{k}{2n}
\label{eqn:cssreduced}
\end{equation}

Here, the compression will be at its worst, being half of the code rate. However, unlike the DG method, this compression
is not dependent on the number of nodes. The second case to consider is when one node is placed in Group 1,
with the rest in Group 2 (i.e. $N_{g_{1}}=1$ and $N_{g_{2}}=N-1$). Also, let Group 1 be allocated no rows (the node will send full information) and Group 2 all the rows (nodes will send the minimum information). In other words, the compression of group 2 is maximised. This is shown in Equation \eqref{eqn:cssbest}:

\begin{equation}
CSS=\frac{kN-k}{Nn}
\label{eqn:cssbest}
\end{equation}

\begin{figure}[H]
\begin{centering}
\includegraphics[width=1\columnwidth]{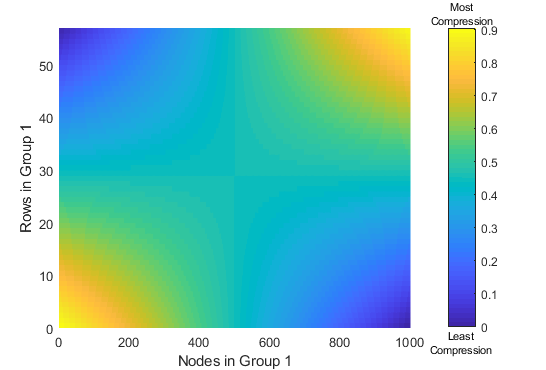}
\par\end{centering}
\caption{Compression ratios of the FG method for different node groupings, using a Hamming(63,57) code and 1000 nodes.}
\label{fig:fgmap}
\end{figure}

If we divide the numerator and denominator by $N$ and let it go to infinity, the second term in the numerator tends to 0, leaving us with $CSS=\frac{k}{n}$. This gives the best compression performance, tending towards the code rate as more nodes are added to the second group. A range of compression performances is illustrated in Figure \ref{fig:fgmap} for various node and row partitions using a Hamming(63,57) and 1000 nodes. The brightest part of Figure \ref{fig:fgmap} is obtained by placing most of the nodes into the full compression group (corresponding to Equation \eqref{eqn:cssreduced}), while the worst compression is derived by doing the inverse. This second arrangement is only the theoretically worst-case scenario however, and does not have any practical benefit. Rather, the worst-case from a practical perspective is in the middle of Figure \ref{fig:fgmap} (i.e. the part calculated with Equation \eqref{eqn:cssbest}). 

To compare the FG and DG methods more fairly, we use 4 nodes, with a Hamming(7,4) ECC. In this case, the DG method will have a $CSS$ of 14.29\%, while the FG method's $CSS$ will be 28.57\% in the most balanced, worst-case scheme and 42.86\% for the high compression scheme. Thus, for this setup, the FG method performs 2-3 times better than the DG method.

\subsection{Lossless Decoding Range}

Both the DG and FG methods share the same LDR.  Using the assumptions in Section \ref{sec:LDR}, the minimum decodable range is equal to the error-correcting capability of the ECC, $t$. This means that all values in the various groups must have a maximum difference less than or equal to $t$. However, as with regular error codes, the scheme will correct a bit in any position. Thus, the scheme might be able to correctly decode values that have a large difference in terms of decimal values, provided that the number of bits that are different is less than $t$.

\subsection{Security and Node Integrity}

The DG method has the most information hiding and data validation, since every node must be decoded with every other
node to decide on the correct value. Thus, a malicious individual would need to control a majority of the nodes in the network in order to alter the information decoded at the server.

The FG method is adaptable, depending on the nodes in a group and the rows assigned to each group. The two cases mentioned in Section \ref{subsec:CSS} are discussed again, this time in terms of the security.
In the case where the rows and nodes are shared equally, all nodes send a slightly compressed version, meaning that there is some information hiding. In the high compression case, one node is sending full information, meaning that there is no security at all. Another factor to note is that one only needs a single node in a particular group to decode all others in the other group. As a result, the overall security is reduced. To mitigate this, the scheme should be set up according to the first case mentioned above and employ a slightly tweaked version of the decoding method. In this altered version, the server performs a grouped decoding method with every combination of node pairs, provided that they are in separate groups, and selects the most likely version using the voting method, which would make the robustness of the data validation more
akin to the DG method. Regardless, the FG method does not perform as well as the DG method in terms of security and node integrity.

\subsection{Network Management Complexity  }
The DG method is reliant on all nodes being present when determining
the row starts and the number of rows. Thus, when a new node joins, the
entire scheme must be redetermined and the new configurations sent
to all nodes. There is also a ceiling to the number of nodes that
can join since a node can have a minimum of one row in the \textbf{$\mathbf{H}^{\mathit{T}}$}
matrix. As a result, for a ($n$, $k$) code, only $k$ nodes can
be in the same scheme. Residual nodes can either be stacked in a particular
group, or an entirely new scheme can be determined for them.

The FG method is designed specifically to minimise network management.
When a new node joins, it simply decides whether the node should join
Group 1 or 2 (according to some predetermined group split percentage)
and assigns the corresponding row start and number of rows to this
node alone. The decoding method will automatically and seamlessly
incorporate the new node into the scheme, meaning the network management
is very much simplified as compared to the DG method. 

\section{Conclusion} \label{sec:conclusion}
Two techniques are presented, in which information can be compressed in a distributed system and losslessly reconstructed at the sink. For both schemes, the limitations and design principles are defined and explained, as well as the practical methods to compress and decompress the information. Of the two techniques, the Flexible Group method has the more adaptable compression, simpler network management of the nodes and is more tolerant to node faults. The lossless decoding range of both schemes are the same and the Disjoint Grouping method (given by literature) has the greater capacity for node validation and information hiding.

\bibliographystyle{IEEEtran}
\bibliography{References}

\begin{thebibliography}{1}
\providecommand{\url}[1]{#1}
\csname url@samestyle\endcsname
\providecommand{\newblock}{\relax}
\providecommand{\bibinfo}[2]{#2}
\providecommand{\BIBentrySTDinterwordspacing}{\spaceskip=0pt\relax}
\providecommand{\BIBentryALTinterwordstretchfactor}{4}
\providecommand{\BIBentryALTinterwordspacing}{\spaceskip=\fontdimen2\font plus
\BIBentryALTinterwordstretchfactor\fontdimen3\font minus
  \fontdimen4\font\relax}
\providecommand{\BIBforeignlanguage}[2]{{%
\expandafter\ifx\csname l@#1\endcsname\relax
\typeout{** WARNING: IEEEtran.bst: No hyphenation pattern has been}%
\typeout{** loaded for the language `#1'. Using the pattern for}%
\typeout{** the default language instead.}%
\else
\language=\csname l@#1\endcsname
\fi
#2}}
\providecommand{\BIBdecl}{\relax}
\BIBdecl

\bibitem{1055037}
D.~Slepian and J.~Wolf, ``Noiseless coding of correlated information sources,''
  \emph{IEEE Transactions on Information Theory}, vol.~19, no.~4, pp. 471--480,
  July 1973.

\bibitem{1281474}
D.~Schonberg, K.~Ramchandran, and S.~S. Pradhan, ``Distributed code
  constructions for the entire slepian-wolf rate region for arbitrarily
  correlated sources,'' in \emph{Data Compression Conference, 2004.
  Proceedings. DCC 2004}, March 2004, pp. 292--301.

\bibitem{755665}
S.~S. Pradhan and K.~Ramchandran, ``Distributed source coding using syndromes
  (discus): design and construction,'' in \emph{Data Compression Conference,
  1999. Proceedings. DCC '99}, Mar 1999, pp. 158--167.

\bibitem{838176}
------, ``Distributed source coding: symmetric rates and applications to sensor
  networks,'' in \emph{Proceedings DCC 2000. Data Compression Conference},
  2000, pp. 363--372.

\bibitem{1328091}
Z.~Xiong, A.~D. Liveris, and S.~Cheng, ``Distributed source coding for sensor
  networks,'' \emph{IEEE Signal Processing Magazine}, vol.~21, no.~5, pp.
  80--94, Sept 2004.

\bibitem{1281475}
V.~Stankovic, A.~D. Liveris, Z.~Xiong, and C.~N. Georghiades, ``Design of
  slepian-wolf codes by channel code partitioning,'' in \emph{Data Compression
  Conference, 2004. Proceedings. DCC 2004}, March 2004, pp. 302--311.

\bibitem{4815077}
S.~Choi, ``Lossy distributed source coding using graphs,'' \emph{IEEE
  Communications Letters}, vol.~13, no.~4, pp. 262--264, April 2009.

\bibitem{1033198}
A.~D. Liveris, Z.~Xiong, and C.~N. Georghiades, ``A distributed source coding
  technique for correlated images using turbo-codes,'' \emph{IEEE
  Communications Letters}, vol.~6, no.~9, pp. 379--381, Sept 2002.

\end{thebibliography}

\end{document}